\documentclass[aps,prb,twocolumn,superscriptaddress,showpacs]{revtex4-1}

\usepackage{graphicx}
\usepackage{calc}
\usepackage{bm}
\usepackage{color}
\usepackage{mathtools}

\begin{document}

\title{Gate-dependent non-linear Hall effect at room temperature in topological semimetal GeTe.}

\author{N.N. Orlova}
\affiliation{Institute of Solid State Physics of the Russian Academy of Sciences, Chernogolovka, Moscow District, 2 Academician Ossipyan str., 142432 Russia}
\author{A.V.~Timonina}
\affiliation{Institute of Solid State Physics of the Russian Academy of Sciences, Chernogolovka, Moscow District, 2 Academician Ossipyan str., 142432 Russia}
\author{N.N.~Kolesnikov}
\affiliation{Institute of Solid State Physics of the Russian Academy of Sciences, Chernogolovka, Moscow District, 2 Academician Ossipyan str., 142432 Russia}
\author{E.V.~Deviatov}
\affiliation{Institute of Solid State Physics of the Russian Academy of Sciences, Chernogolovka, Moscow District, 2 Academician Ossipyan str., 142432 Russia}

\date{\today}

\begin{abstract}
We experimentally investigate non-linear Hall effect as  zero-frequency and  second-harmonic transverse voltage responses to ac electric current for  topological semimetal GeTe. A thick single-crystal GeTe flake is placed on the Si/SiO$_2$ substrate, where the p-doped Si layer serves as a gate electrode. We confirm, that electron concentration is not gate-sensitive in thick GeTe flakes due to the gate field screening by bulk carriers. In contrast, by transverse voltage measurements, we demonstrate that the non-linear Hall effect shows pronounced dependence on the gate electric field at room temperature. Since the non-linear Hall effect is a direct consequence of a Berry curvature dipole in topological media, our observations indicate that Berry curvature can be controlled by the  gate electric field. This experimental observation can be understood as a result of the known dependence of giant Rashba splitting on the external electric field in GeTe. For possible applications,  the zero-frequency gate-controlled non-linear Hall effect can be used for the efficient broad-band rectification.
\end{abstract}

\maketitle

\section{Introduction}

Recent interest to the non-linear Hall (NLH) effect~\cite{sodemann}  is connected  with its significance both for the fundamental physics and for possible applications. In the linear response, there is no Hall voltage in the presence of time-reversal symmetry.  NLH effect is  predicted~\cite{sodemann} as a transverse voltage response in zero magnetic field due to the Berry curvature dipole in momentum   space~\cite{deyo,golub,moore,low,nlhe1,nlhe2,nlhe3,nlhe4,nlhe5,nlhe6}. Thus, for the fundamental physics, NLH effect is the direct manifestation of finite Berry curvature in topological media.  Since Berry curvature often concentrates in regions  where two or more bands cross, three classes of candidate materials have been proposed~\cite{sodemann}: topological crystalline insulators, two-dimensional transition metal dichalcogenides, and three-dimensional Weyl and Dirac semimetals.

Being defined by the bulk energy spectrum, NLH effect is a rare example of macroscopic quantum phenomenon, which does not obligatory require low temperatures to be observed. A search for the room-temperature effect points to three-dimensional systems, primary topological semimetals~\cite{armitage}. In Weyl semimetals every band touching point splits  into two Weyl nodes with opposite chiralities due to the time reversal or inversion symmetries breaking. The materials with broken time-reversal symmetry are bulk ferromagnets or antiferromagnets, while Weyl semimetals with broken inversion symmetry have to obtain bulk ferroelectric polarization~\cite{armitage,TSreview}. Due to  the gapless bulk spectrum~\cite{armitage}, ferroelectric polarization makes  the non-magnetic Weyl semimetal being the natural  representation of the  novel concept of the  intrinsic polar metal, or the ferroelectric conductor~\cite{PM,pm1,pm2,pm4}. 

As a transverse response to longitudinal ac excitation, NLH effect can be observed both at zero and twice the excitation frequency~\cite{golub}. The second-harmonic response is much easier to be measured by standard lockin technique~\cite{ma,kang,esin,c_axis}. On the other hand, zero-frequency  NLH response can be used for high-frequency (even terahertz or infrared) detection~\cite{NLHErect,NLHErect1}, which is important for wide-band communications~\cite{tera},  wireless charging, energy harvesting, etc. An advantage of the NLH rectification is the absence of thermal losses, since it originates from the Berry curvature dipole. The latter can be in principle controlled by electric field, which has been demonstrated~\cite{NLHEgate} for two-dimensional WTe$_2$. 

Thus, both physics and applications require new materials for the room-temperature NLH effect, which  allow electric field control of Berry curvature dipole. Among these materials, GeTe is of special interest~\cite{GeTespin-to-charge,GeTereview} due to the reported  giant Rashba splitting~\cite{GeTerashba,GeTerashba1}. GeTe is predicted to be topological semimetal in low-temperature ferroelectric $\alpha$-phase~\cite{ortix,triple-point}.   As an additional advantage, 
the Rashba parameter is known to depend on the external electric field in GeTe~\cite{GeTerashba,spin text,GeTeour}.

Here, we experimentally investigate non-linear Hall effect as  zero-frequency and  second-harmonic transverse voltage responses to ac electric current for  topological semimetal GeTe. We confirm, that electron concentration is not gate-sensitive in thick conductive flakes due to the gate field screening by bulk carriers in GeTe. In contrast, by transverse voltage measurements, we demonstrate that the non-linear Hall effect shows pronounced dependence on the gate electric field at room temperature.

\section{Samples and technique}

\begin{figure}
\includegraphics[width=1\columnwidth]{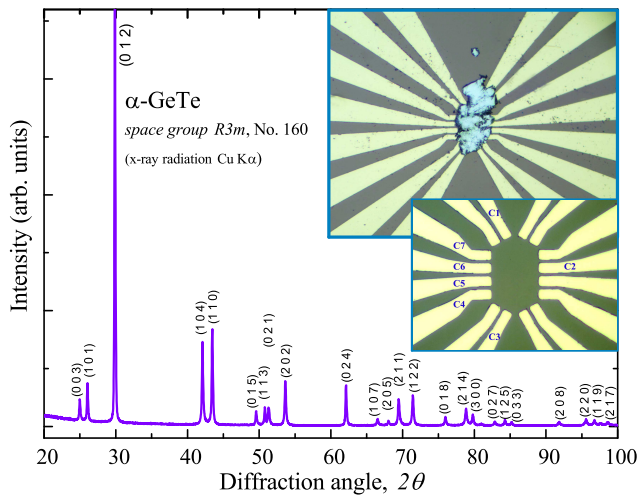}
\caption{(Color online) (a) The X-ray powder diffraction  pattern (Cu K$\alpha$ radiation), which is obtained for the crushed GeTe single crystal. The single-phase  $\alpha$-GeTe is confirmed with the space group R3m (No.160).
The upper inset shows optical image of the sample with a 0.5~$\mu$m thick GeTe flake on the insulating SiO$_2$ substrate. The Au leads geometry is shown below. 100 nm thick, 5~$\mu\mbox{m}$ separated  leads form  Hall-bar geometry with several pairs of current and potential contacts. The ac current is applied between C1 and C3 leads (80~$\mu$m distance), while the transverse (Hall) voltage $U_{xy}$ is measured between the  C2 and C6 potential probes (60~$\mu$m). Also, the longitudinal $U_{xx}$ component can be measured between the C6 and C5 (5~$\mu$m separation). The p-doped bulk Si layer under the SiO$_2$ surface serves as a gate electrode to apply the gate electric field  through the 200~nm thick SiO$_2$ layer.
  }
\label{sample}
\end{figure}

GeTe single crystals were grown by physical vapor transport in the evacuated silica ampule. The initial GeTe load was synthesized by direct reaction of the high-purity (99.9999\%) elements in vacuum. For the  crystals growth, the initial GeTe load serves as a source of vapors: it was melted and kept at 770-780$^\circ$ C for 24 h. Afterward, the source was cooled down to 350$^\circ$ C at the 7.5 deg/h rate. The GeTe crystals grew during this process on the cold  ampule walls somewhat above the source. 

The GeTe composition is verified by energy-dispersive X-ray spectroscopy. The powder X-ray diffraction analysis confirms single-phase GeTe, see Fig.~\ref{sample} (a),  the known structure model~\cite{GeTerashba} is also refined with single crystal X-ray diffraction measurements. 

Ferroelectric polarization was previously reported for the epitaxal films, microwires and bulk GeTe crystals~\cite{GeTespin-to-charge}, it is defined by the non-centrosymmetric distorted rhombohedral structure ($\alpha-GeTe$) with space group R3m (No. 160)~\cite{GeTerashba}. 
The gate-dependent ferroelectric polarization was also demonstrated in capacitance measurements~\cite{GeTeour}, the results confirmed giant Rashba splitting~\cite{GeTerashba} in our GeTe single crystals. 

\begin{figure}
\includegraphics[width=\columnwidth]{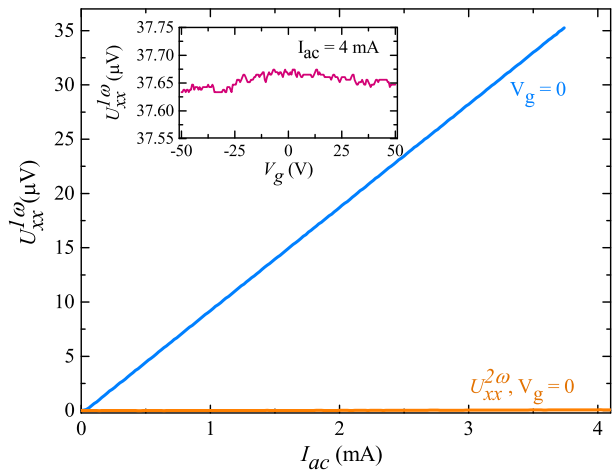}
\caption{(Color online) Standard Ohmic behavior for the first-harmonic  $U^{1\omega}_{xx}$ longitudinal voltage component. $U^{1\omega}_{xx}(I_{ac})$ is strictly linear, so the second-harmonic $U^{2\omega}_{xx}$ component is  neglegibly small. The measured $U^{1\omega}_{xx}(I_{ac})$ slope corresponds to the $\approx 10$~m$\Omega$ GeTe resistance between the C6 and C5 probes in Fig.~\ref{sample}. Inset shows  the first-harmonic  $U^{1\omega}_{xx}$ responce independence of the gate voltage in a wide $\pm 50$~V range, as it should be expected for the gate field screening in thick conductive GeTe flakes.}
\label{VACxx}
\end{figure}

The upper inset to Fig.~\ref{sample} shows a top-view image of the sample.  The topological semimetals are essentially three-dimensional objects~\cite{armitage}, so we have to select  relatively thick (above 0.5~$\mu$m) flakes. A thick flake  also ensures sample homogeneity for correct determination of xx- and xy- voltage responses, however, the desired experimental geometry can not be defined by usual mesa etching for thick flakes. 

Thick flakes require special sample preparation procedure: the mechanically exfoliated GeTe flake is transferred on the Au leads pattern, which is defined by lift-off technique on the SiO$_2$ surface after thermal evaporation of 100~nm Au, as depicted in the inset to Fig.~\ref{sample}. We choose the $\approx$100~$\mu$m wide flakes with defect-free surface by optical microscope. After initial single-shot pressing by another oxidized silicon substrate,  the flake is firmly connected to the Au leads. This procedure provides high-quality contacts to the flake, while the Au leads pattern defines the desired experimental geometry. The obtained samples are electrically and mechanically  stable even in different cooling cycles to liquid helium temperatures.   This sample preparation technique has been verified  before for a wide range of materials~\cite{cdas,cosns,inwte2,incosns,timnal,black,infgt,aunite}.  

We investigate  transverse (xy-)  first- and second-harmonic  voltage responses by standard four-point lock-in technique.  The ac current is applied between C1 and C3 contacts in Fig.~\ref{sample} (b), while the transverse (Hall) voltage $U_{xy}$   is measured between the  C2 and C6 potential probes. Also, the longitudinal $U_{xx}$ component can be measured between the C6 and C5. The Au leads are of 10~$\mu$m width, the current ones C1 and C3 are separated by 80~$\mu$m distance. The voltage probe  separation is 60~$\mu$m for the xy- configuration (C2 and C6) and 5~$\mu$m for the xx- one (C6 and C5), respectively.  The signal is confirmed to be independent of the ac current frequency within 100 Hz -- 10kHz range, which is defined by the applied filters.

The p-doped bulk Si layer under the SiO$_2$ surface serves as a gate electrode to apply the gate electric field  through the 200~nm thick SiO$_2$ layer. Even for relatively thick flakes,  ferroelectric polarization is sensitive~\cite{WTeour,GeTeour} to the gate electric field, since the relevant (bottom) flake surface is directly adjoined to the SiO$_2$ layer. We check by electrometer that there is no measurable leakage current in the gate voltage range  $\pm 50$~V. Since the ferroelectric $\alpha-GeTe$  phase exists~\cite{Tc} below $700$~K, and  the most important GeTe surface with Au contacts (the bottom one) is protected from any contamination by SiO$_2$ substrate,  all the measurements are performed at room temperature under ambient conditions. This might be also important in the case of possible applications of the observed effects.

\section{Experimental results}

We confirm the correctness of  experimental conditions by demonstrating standard Ohmic behavior for the first-harmonic  $U^{1\omega}_{xx}$ longitudinal voltage component, see  Fig.~\ref{VACxx}.   $U^{1\omega}_{xx}(I_{ac})$ shows strictly linear dependence on the applied ac current $I_{ac}$, which is also confirmed by negligibly small second-harmonic  $U^{2\omega}_{xx}$ response. The measured $U^{1\omega}_{xx}(I_{ac})$ slope corresponds to the $\approx 10$~m$\Omega$ sample resistance  between the C6 and C5 probes in Fig.~\ref{sample}. 

\begin{figure}
\includegraphics[width=\columnwidth]{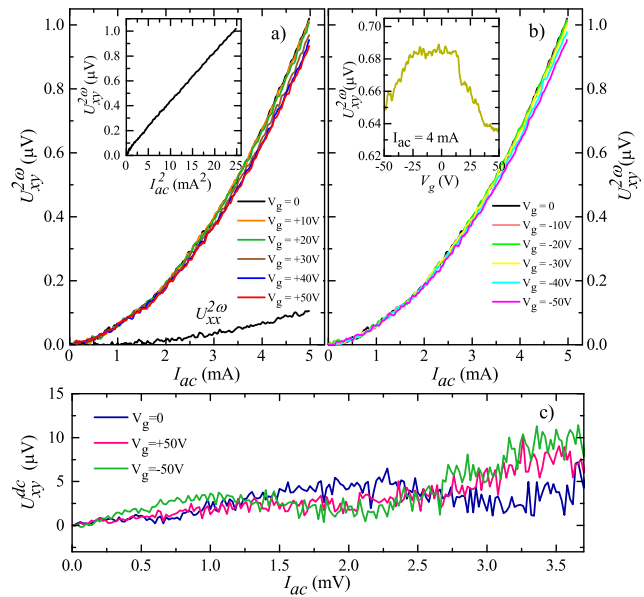}
\caption{(Color online) (a) and (b) Typical NLH behavior~\cite{ma,kang,esin,c_axis} of the transverse  second-harmonic voltage component at different gate voltages.  $U^{2\omega}_{xy}\sim I^2$ is confirmed in the inset to (a) as strictly linear $U^{2\omega}_{xy}(I^2)$ dependence for zero gate voltage. The longitudinal  second-harmonic voltage $U^{2\omega}_{xx}$  is one order of magnitude smaller. The curves are shown for positive (a) and negative (b) gate voltages, they are symmetrically affected by the gate electric field of both signs. Inset to (b) demonstrates the gate voltage scan $U^{2\omega}_{xy}(V_g)$ at  fixed ac current $I_{ac}=4$~mA, $U^{2\omega}_{xy}$ is symmetrically diminishing within $\approx 10$~\%.  (c) Zero-frequency transverse  voltage component $U^{dc}_{xy}$ as function of the applied ac current $I_{ac}$ at three fixed gate voltages $V_g=0, \pm50$~V. The curves are strongly nonlinear, $U^{dc}_{xy}$ also symmetrically depends on the gate voltage of both signs.  The data are obtained at room temperature.}
\label{VAC}
\end{figure}

In contrast to the standard two-dimensional materials like graphene or field-effect transistors, carrier concentration in thick three-dimensional GeTe flakes is not sensitive to the gate electric field due to the perfect field screening by bulk carriers.  This is experimentally confirmed in the inset to Fig.~\ref{VACxx}:  the longitudinal $U^{2\omega}_{xx}$ response is nearly independent of the gate voltage with 0.1~\% accuracy in a wide gate voltage range $\pm50$~V. Maximum variation of the carrier concentration can be estimated in the capacitor approximation as $\delta n / n \sim \delta R /R \approx 10^{-3} $. 

Fig.~\ref{VAC}  shows typical behavior of the non-linear Hall effect~\cite{ma,kang,esin,c_axis}  as a quadratic transverse Hall-like response $U^{2\omega}_{xy}$ to ac excitation current $I$. The $\sim I^2$ dependence  is directly demonstrated in the inset to Fig.~\ref{VAC} (a) for zero gate voltage. The longitudinal  second-harmonic voltage $U^{2\omega}_{xx}$ is one order of magnitude smaller, which confirms well-defined Au leads geometry and high sample homogeneity.  

The non-linear Hall $U^{2\omega}_{xy}$ curves show clearly visible dependence on the gate voltage in Fig.~\ref{VAC} (a) and (b). For the whole current range,  $U^{2\omega}_{xy}$ values are diminishing both for positive gate voltages in Fig.~\ref{VAC} (a), and for the negative ones in Fig.~\ref{VAC} (b). This behavior is directly demonstrated in the inset to Fig.~\ref{VAC} (b) as the gate voltage scan $U^{2\omega}_{xy}(V_g)$ at the fixed current value $I_{ac}=4$~mV. The  scan indeed shows symmetrical  $U^{2\omega}_{xy}$ diminishing within $\approx 10$\% for both gate voltage signs, while the $U^{2\omega}_{xy}(V_g)$ curve is centered at $V_g=-10$~V. It is important, that this symmetric $U^{2\omega}_{xy}(V_g)$ dependence can not be ascribed to the gate-field effect on the carrier concentration in GeTe: (i) there is no noticeable dependence in the inset to Fig.~\ref{VACxx}; (ii) the concentration should depend asymmetrically on the gate voltage sign.   

If one demonstrates NLH effect at twice the frequency of the excitation current, it is natural to expect NLH  rectification as the zero-frequency dc voltage $U^{dc}_{xy}$. Indeed, we observe finite $U^{dc}_{xy}$ values for the applied ac current $I_{ac}$ in Fig.~\ref{VAC} (c).  $U^{dc}_{xy}(I_{ac})$ is clearly non-linear, while it does not show clear $\sim I^2$ dependence in a whole ac current range. The dc signal is noisy in comparison with the  second-harmonic one due to the direct measurements by digital dc voltmeter after a broad-band preamplifier. Direct measurements of low signals by a voltmeter might be also a reason  to have distorted non-linear $U^{dc}_{xy}(I_{ac})$ dependence in comparison with the clear $\sim I^2$ signal from lockin. In both the cases, we check the signal to be antisymmetric if one exchanges the Hall voltage probes. Also, $U^{dc}_{xy}(I_{ac})$ shows symmetrical gate voltage dependence in Fig.~\ref{VAC} (c), despite the dc voltage is increasing for both signs of $V_g$. As a result, we observe finite gate-dependent non-linear transverse dc voltage $U^{dc}_{xy}(I_{ac})$, which confirms NLH rectification in GeTe at room temperature. 

\begin{figure}
\includegraphics[width=\columnwidth]{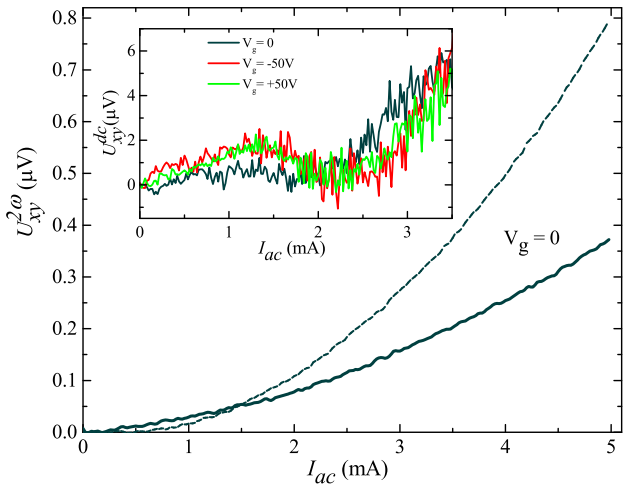}
\caption{(Color online) Room-temperature NLH behavior of the transverse  second-harmonic voltage component $U^{2\omega}_{xy}\sim I^2$ for two different samples, as depicted by solid (60~$\mu$m spacing between the voltage probes) and dashed (20~$\mu$m spacing) lines, respectively. The curves are obtained at zero gate voltage. Inset shows the gate voltage dependence of the zero-frequency transverse  voltage component $U^{dc}_{xy}$ for one of the samples (solid curve in the main field). The curves are strongly nonlinear, $U^{dc}_{xy}$ also symmetrically depends on the gate voltage of both signs, similarly to Fig.~\ref{VAC} (c).  }
\label{VACsmall}
\end{figure}

Qualitatively similar $\sim I^2$  NLH dependence  can be demonstrated for different samples with different distances between the voltage probes, see Fig.~\ref{VACsmall}, as depicted by solid and dashed lines, respectively. The curves are quite similar for the samples of the same dimensions, see the dashed  curve in Fig.~\ref{VACsmall} and the curves in Fig.~\ref{VAC}. For the 20~$\mu$m spaced voltage leads (solid line in Fig.~\ref{VACsmall}),  the $U^{2\omega}_{xy}$ values are twice smaller in comparison with Fig.~\ref{VAC} (a).  The zero-frequency $U^{dc}_{xy}(I_{ac})$ curves are presented in the inset to Fig.~\ref{VACsmall} for the smallest (20~$\mu$m) sample. The curves are qualitatively similar to ones in Fig.~\ref{VAC} (c): the obtained values  are twice smaller, the curves are strongly nonlinear, $U^{dc}_{xy}$ also symmetrically depends on the gate voltage for both signs of $V_g$.

\section{Discussion} \label{disc}

As a result, we observe second-harmonic NLH effect and NLH rectification at room temperatures in GeTe topological semimetal, both effects are sensitive to the gate electric field. 

The effect of the gate voltage is about 10~\% in Fig.~\ref{VAC}. However, it is not only important to control the observed $U^{2\omega}_{xy}$ and $U^{dc}_{xy}$, but also allows to rule out possible influence of thermopower on the transverse voltage response. In principle, topological materials are  characterized by strong thermoelectricity~\cite{ptsn,cdas_thermo}, which also appears as a second-harmonic signal~\cite{kvon,shashkin}. For an inhomogeneous sample, it is possible to expect that Hall voltage probes are not perfectly symmetric in respect to the current path. In this case, Joule heating $\sim R I^2$ can produce temperature gradient between the Hall voltage probes, so $U_{xy} \sim R I^2$ could be expected from thermoelectricity~\cite{avci,esin}. However, the sample resistance $R=U^{1\omega}_{xx}/I_{ac}$ does not depend on the gate voltage in the inset to Fig.~\ref{VACxx}.  In contrast,  $U^{2\omega}_{xy}(V_g)$ shows the symmetric 10~\% variation in Fig.~\ref{VAC}, so  the observed $U^{2\omega}_{xy}(V_g)$ behavior is not connected with thermoelectricity~\cite{esin}. 

The non-linear Hall effect~\cite{ma,kang,esin,c_axis} arises from  the Berry curvature dipole in momentum space~\cite{sodemann}. In the simplified picture, the longitudinal current generates the effective sample magnetization, which leads to the Hall effect even in zero external magnetic field. Hall voltage is therefore  proportional to the square of the excitation current, so it can be detected as the second-harmonic  $U^{2\omega}_{xy}$ or the zero-frequency $U^{dc}_{xy}$ transverse voltage components, as we observe in Figs.~\ref{VAC} and~\ref{VACsmall}. In this simple model it is obvious that the Berry curvature dipole can be controlled by dc electric field, as it has been demonstrated~\cite{NLHEgate} for two-dimensional WTe$_2$:  in-plane dc electric field $E_{in}$ affects the measured second-harmonic NLH signal $U^{2\omega}_{xy}$ if the field $E_{in}$ is parallel to the ac excitation current $I_{ac}$. This argumentation can not be applied to our experimental conditions, where the NLH voltage is affected by out-of-plane gate electric field $E_g$ in Figs.~\ref{VAC} and~\ref{VACsmall}.

Since the carrier concentration is independent of the gate electric field in our samples, see the inset to Fig.~\ref{VACxx}, it is the Rashba parameter $\alpha_R$ which defines the measured NLH signal $U^{2\omega}_{xy}$. In non-magnetic topological semimetals, the Berry curvature is defined by Weyl nodes separation due to the breaking of inversion symmetry~\cite{armitage,TSreview}, so the Berry curvature is directly connected with the Rashba parameter $\alpha_R$ in GeTe. On the other hand, the dependence of the Rashba parameter on the ferroelectric polarization is known for giant Rashba splitting in GeTe from theoretical~\cite{GeTerashba} and experimental~\cite{spin text,GeTeour} investigations. Even for relatively thick flakes,  ferroelectric polarization is sensitive~\cite{WTeour,GeTeour} to the gate electric field, since the relevant (bottom) flake surface is directly adjoined to the SiO$_2$ layer. Variation of $\alpha_R$ can be as large~\cite{GeTerashba,GeTeour} as 20~\% in the gate voltage range of Fig.~\ref{VAC}.  The smaller value of the effect (about 10~\% in Fig.~\ref{VAC}) seems to be because $\alpha_R$ is changed only at the bottom (adjoined to the SiO$_2$ layer) flake surface. Thus, the measured NLH signal $U^{2\omega}_{xy}$ is affected by gate electric field through the field dependence of the giant Rashba spin-orbit coupling in GeTe~\cite{GeTerashba,spin text,GeTeour}.

\section{Conclusion}
As a conclusion, we observe second-harmonic NLH effect and NLH rectification at room temperatures in GeTe topological semimetal, both effects are sensitive to the gate electric field. The observed behavior is important both for the fundamental physics and for possible applications. For physics, NLH effect is a direct consequence of a Berry curvature dipole, which we demonstrate to be controlled by the  gate electric field in GeTe. For possible applications,  the gate-controlled NLH rectification can be used for the efficient broad-band detection.

\acknowledgments

We wish to thank S.S~Khasanov for X-ray sample characterization and A.A.~Avakyants and D.Yu.~Kazmin for the help in sample preparation.  We gratefully acknowledge financial support  by the  Russian Science Foundation, project RSF-23-22-00142, https://rscf.ru/project/23-22-00142/.

\end{document}